\documentclass[twocolumn]{aastex63}
\usepackage[utf8]{inputenc}

\def\flunits{${\rm erg\; cm^{-2}\; s^{-1}\; \AA^{-1}}$}

\begin{document}

\title{Peeking Between the Pulses: The Far-UV Spectrum\\ of the Previously Unseen White Dwarf in AR~Scorpii}

\author{Peter Garnavich}
\affiliation{University of Notre Dame, Notre Dame, IN 46556, USA}
\author{Colin Littlefield}
\affiliation{University of Notre Dame, Notre Dame, IN 46556, USA}
\author{Maxim Lyutikov}
\affiliation{Purdue University, West Lafayette, IN 47907, USA}
\author{Maxim Barkov}
\affiliation{Purdue University, West Lafayette, IN 47907, USA}
\affiliation{RIKEN, Wako, Saitama 351-0198, Japan}
\affiliation{Institute of Astronomy, Russian Academy of Sciences, Moscow, 119017 Russia}

\accepted{for publication in the Astrophysical Jorunal, December 16, 2020 }

\begin{abstract}

The compact object in the interacting binary AR Sco has widely been presumed to be a rapidly rotating, magnetized white dwarf (WD), but it has never been detected directly. Isolating its spectrum has proven difficult because the spin-down of the WD generates pulsed synchrotron radiation that far outshines the WD's photosphere. As a result, a previous study of AR~Sco was unable to detect the WD in the averaged far-ultraviolet spectrum from a Hubble Space Telescope ({\it HST}) observation. In an effort to unveil the WD's spectrum, we reanalyze these {\it HST} observations by calculating the average spectrum in the troughs between synchrotron pulses. We identify weak spectral features from the previously unseen WD and estimate its surface temperature to be 11500$\pm$500~K. Additionally, during the synchrotron pulses, we detect broad Lyman-$\alpha$ absorption consistent with hot WD spectral models. We infer the presence of a pair of hotspots, with temperatures between 23000~K and 28000~K, near the magnetic poles of the WD. As the WD is not expected to be accreting from its companion, we describe two possible mechanisms for heating the magnetic poles. The Lyman-$\alpha$ absorption of the hotspots appears relatively undistorted by Zeeman splitting, constraining the WD's field strength to be $\lesssim$100~MG, but the data are insufficient to search for the subtle Zeeman splits expected at lower field strengths.

\end{abstract}


\section{Introduction}

The unique interacting binary system AR Scorpii (AR~Sco hereafter) has been called a white dwarf ``pulsar'' because its bright electromagnetic flashes appear to be generated by the spin-down energy of its degenerate primary stellar component \citep{marsh16, stiller18, gaibor20}. The binary consists of a low-mass red dwarf star and a rapidly spinning magnetized white dwarf (WD) orbiting over a 3.56-hour period. The spectacular pulsed emission is seen over a broad range of wavelengths from radio to soft X-rays \citep{marsh16, takata18, marcote17, stanway} and it is related to the 1.95-minute spin period of the WD and its magnetic interaction with the red dwarf secondary. The pulsed emission does not appear to be dominated by accretion onto the WD \citep{marsh16} as is seen in intermediate polar-type (IP) cataclysmic variable stars (CVs). A study of the polarized emission by \citet{buckley17} showed that the pulses are consistent with synchrotron radiation coming from near the WD, although where and how the electrons are accelerated remains an interesting question.

There have been several models proposed to explain the observed synchrotron pulses. For example, a direct interaction between the WD magnetic field with the secondary star or its wind has been suggested \citep{geng16, katz17}. \citet{garnavich19} identified slingshot prominences from the red dwarf star and speculated that magnetic reconnection events between the WD and secondary star fields are the source of the energetic electrons. These models require extremely high WD field strengths of more that 200~MG to generate sufficient interaction energy near the surface of the red dwarf. 

\begin{figure}[h!]
    \centering
    \includegraphics[width=\columnwidth]{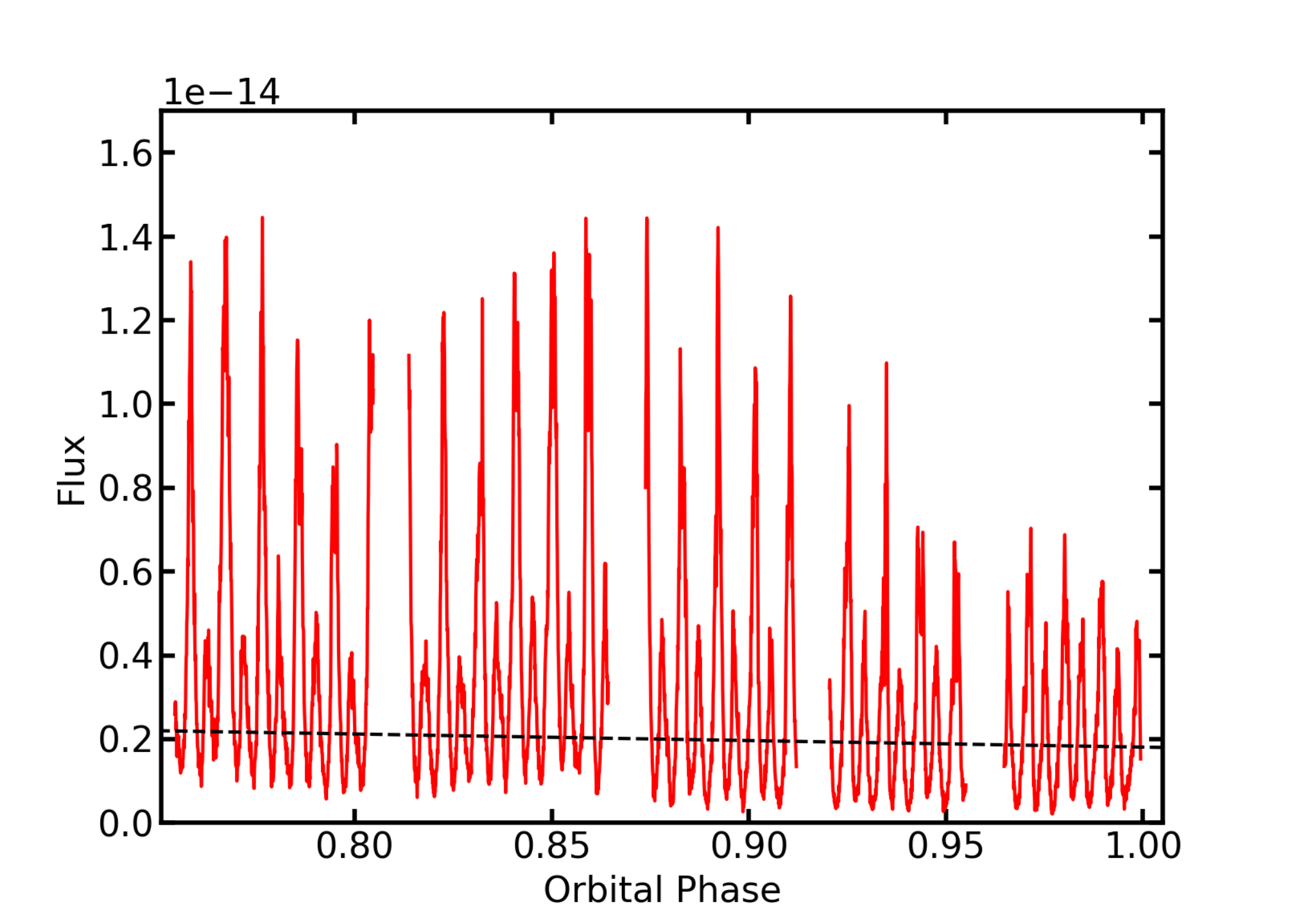}
    \caption{A sample of the $COS/HST$ light curve of AR~Sco showing five individual exposures (obtained over two binary orbits). The dashed line indicates where the division between pulsed photons and trough photons were made in re-sampling the spectra.}
    \label{lc}
\end{figure}

Recently, \citet{lyutikov20} has argued that the surface magnetic field of the WD in AR~Sco must be only about 10~MG to have been able to spin-up during a period of rapid mass accretion. This field strength is similar to what is seen in IPs and that AR~Sco may be an IP in a propeller mode where the spinning WD field drives away the gas donated by the red dwarf. AE~Aqr is the only confirmed case of an IP in a propeller state, but it does not generate synchrotron pulses at the spin period of its WD.   

The \citet{lyutikov20} model places the site of the interaction well inside the WD magnetosphere by allowing mostly neutral gas from the secondary to fall toward the WD before it is ionized and swept away by the spinning magnetic field. They estimate that the  temperature of the WD needed to photoionize the gas is at least 12000$^\circ$K. \citet{marsh16} could not directly see evidence of the WD in the AR~Sco spectra, but placed a rough limit of 9750$^\circ$K on WD surface temperature.

Here, we will attempt to test the \citet{lyutikov20} model by better constraining the WD surface temperature.  We analyze the archival AR~Sco spectra obtained in 2016 by the Hubble Space Telescope's ($HST$) Cosmic Origins Spectrograph ($COS$) with the goal separating the pulsed emission from the inter-pulse light. Reducing the synchroton background by isolating the emission between pulse may reveal the presence of the WD.

\begin{figure}
    \centering
    \includegraphics[width=\columnwidth]{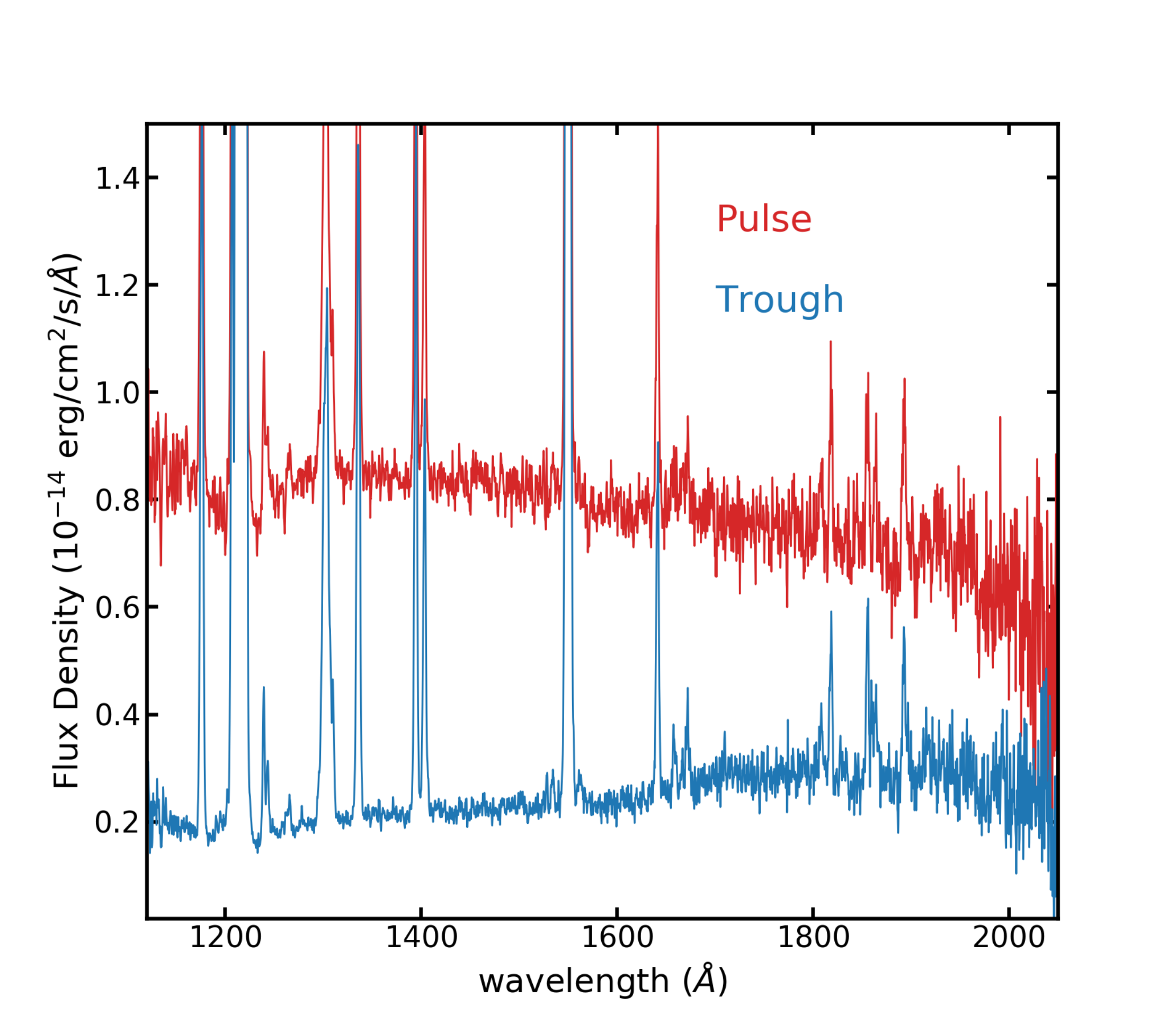}
    \caption{Comparison between the $COS/HST$ spectra obtained during the pulses and the light curve minima (troughs). The pulsed spectrum shows a blue continuum while the trough spectrum is rather red. The emission lines are thought to arise from the irradiated face of the red dwarf and are similar in strength for both subsets of data.}
    \label{pulse}
\end{figure}

\section{Data}

The far-ultraviolet (FUV) region of the spectrum is a good place to search for emission from the WD surface in cataclysmic variable stars \citep[e.g.][]{gaensicke99}. The FUV ranges between 1100~\AA\ and about 2000~\AA\ and contains the Lyman-$\alpha$ absorption feature that can be easy to spot in DA type WDs. In AR~Sco, however, there are other sources of emission that may interfere in characterizing its WD. For example, the $COS/HST$ FUV photometry published by \citet{marsh16} shows strong synchrotron pulses, just as observed in the optical (see Figure~\ref{lc}), as well as a brightness modulation over an orbit. 

To reduce the contribution of the pulses to the FUV spectrum, we have reanalyzed the $COS/HST$ data for AR~Sco by selecting time-tagged photons that arrived between pulses. We call the period around the minimum between pulses the ``trough''. Using the routine {\bf splittag}\footnote{https://justincely.github.io/AAS224/splittag\_tutorial.html}, we divide the 16 {\bf corrtag} exposures in the dataset into pulse-dominated phases and trough-dominated ones with the division occurring at approximately 2$\times 10^{-15}$ \flunits\ as shown in Figure~\ref{lc}. It is not possible to set a single flux value for the division between pulse and trough because, as in the optical, there is an orbital modulation in the FUV light curve. The orbital modulation \citep[e.g.,][]{gaibor20} has an amplitude of more than a magnitude at optical and FUV wavelengths and it peaks around orbital phases\footnote{Zero orbital phase, $\phi$, is defined as inferior conjunction when the red dwarf is between the Earth and the WD} of 0.4 to 0.5.

The time-tag photons were divided into 172 pulses (both primary and secondary pulses were included) and 169 troughs. These sub-exposures are typically between 20 seconds and 40 seconds in length. Each {\bf corrtag} section was converted to an extracted spectrum using the {\bf x1dcorr} routine. The individual pulsed spectra were combined using the {\bf splice}
\footnote{https://www.stsci.edu/itt/review/dhb\_
2013/COS/ch5\_cos\_analysis4.html} routine and the same was done for all the trough spectral pieces. The resulting average spectra are shown in Figure~\ref{pulse}. The emission lines are thought to come from the heated face of the secondary star and are of similar strength in both spectra. The pulsed spectrum has a blue continuum while the trough spectrum increases into longer wavelengths.

\section{Analysis}

\subsection{The Trough Spectrum}
    
\begin{figure}
    \centering
    \includegraphics[width=\columnwidth]{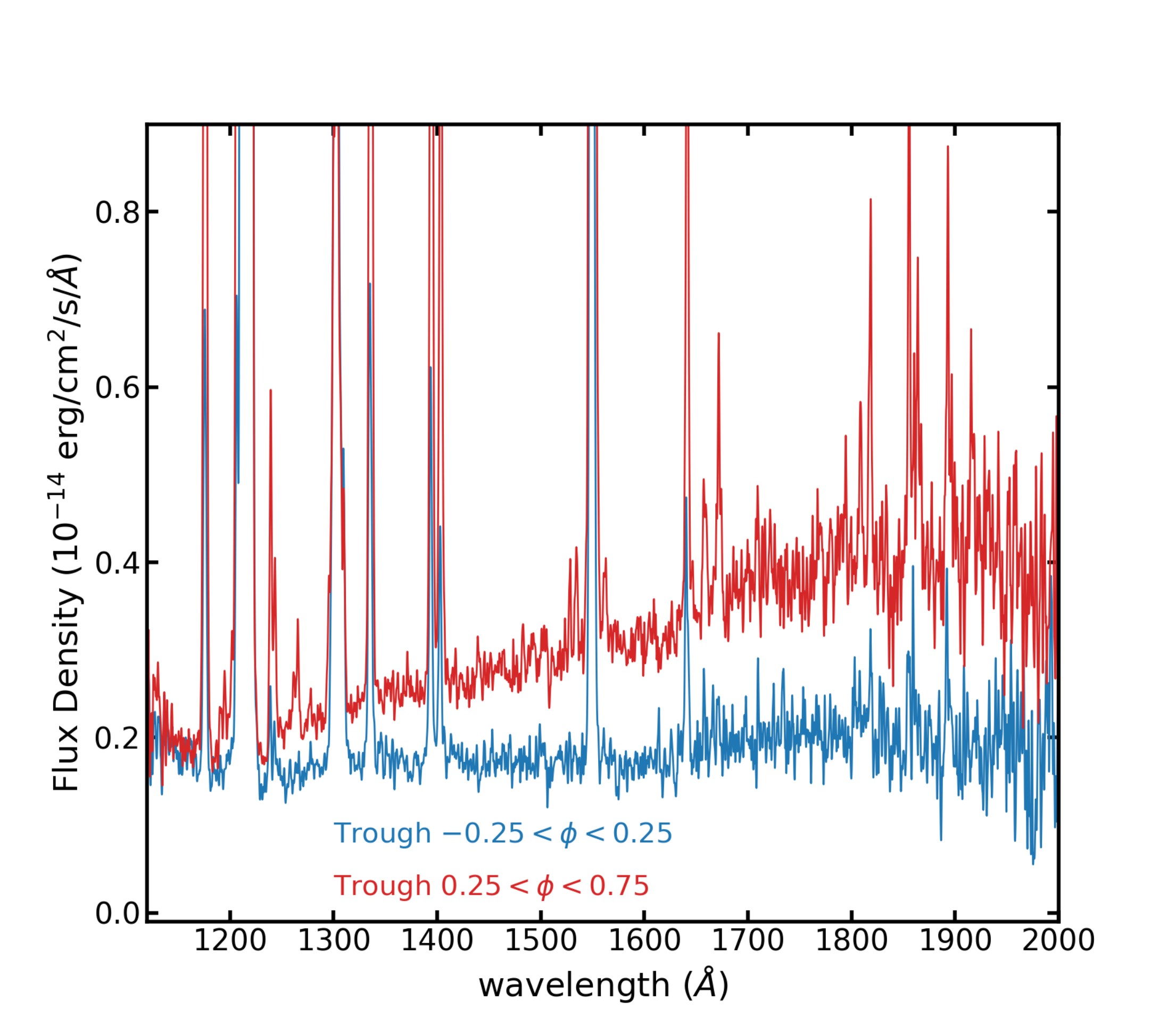}
    \caption{Comparison between trough spectra taken around inferior and superior conjunction. For the spectra taken near superior conjunction the there is a red continuum peaking at approximately 1900~\AA , corresponding to a black body temperature of 15000$^\circ$K. The emission lines are also significantly brighter implying they originate near the irradiated face of the secondary. The spectrum around inferior conjunction is relatively flat. }
    \label{orbit}
\end{figure}

The spectrum of the troughs shown in Figure~\ref{pulse} should provide some limits on the WD properties. However, the orbital light curve modulation seen in the optical also contributes to the FUV light. The orbital variation is likely linked to the heated face of the secondary star. At inferior conjunction the the heated face is partly obscured by the secondary itself, while a half orbit later the emission is seen unimpeded.  To determine how large this contribution is in the FUV, we further sub-divided the trough spectra by orbital phase. The peak of the orbital modulation occurs between orbital phase $0.4< \phi <0.5$. We re-binned the trough spectra into orbital phases with $-0.25 < \phi <0.25$ (centered on inferior conjunction) and phases $0.25 < \phi <0.75$ (centered on superior conjunction). The resulting spectra are shown in Figure~\ref{orbit}.

\begin{figure}
    \centering
    \includegraphics[width=\columnwidth]{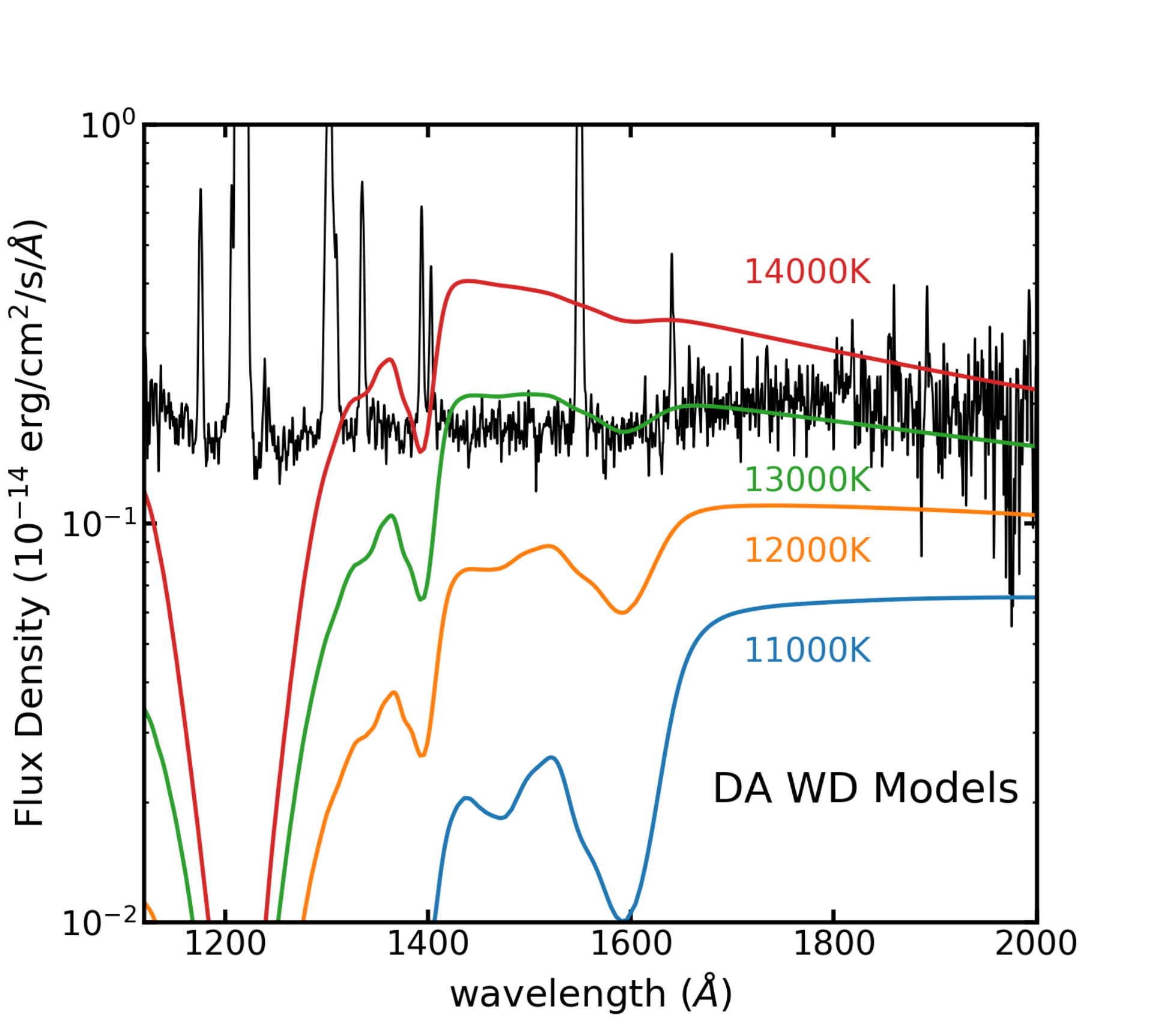}
    \caption{The trough spectrum around inferior conjunction compared with hydrogen-rich WD spectra from \citet{koester10} ranging in surface temperature from 11000$^\circ$K to 14000$^\circ$K. All the models assume $log(g)=8.5$ and a mass for the WD of 0.8~M$_\odot$. The models have been scaled to the distance of AR~Sco. The lack of strong features in the observed spectrum shortward of 1700~\AA\ suggests that a single temperature hydrogen-rich WD is not a good match to the data. Additional sources of emission are required. But DA models with temperatures greater than 13000$^\circ$K are excluded by this analysis.}
    \label{temperature}
\end{figure}

The trough spectrum obtained around superior conjunction clearly shows more flux at the longer wavelengths than at the spectrum around inferior conjunction. We attribute this emission to our better view of the irradiated face of the secondary star during superior conjunction. The spectrum appears to peak around 1900~\AA\  corresponding to a black body temperature of approximately 15000$^\circ$K. 

The trough spectrum around inferior conjunction contains the minimum contamination from the synchrotron pulses and heated face of the secondary star, and therefore it should provide the best constraint on the properties of the WD in the system. We compare the spectrum with \citet{koester10} models assuming a WD mass of 0.8~M$_\odot$ and $logg=8.5$, scaled to the distance of AR~Sco. The choice of $logg=8.5$ for this mass WD is consistent with the recent study by \citet{chandra20}. Different choices for the mass or gravity are possible as these are not well constrained. 

Further, a significant surface magnetic field will reduce the apparent strength of absorption features due to Zeeman splitting. For example, the Lyman-$\alpha$ feature for a DA WD with surface magnetic fields over 100~MG can divide into multiple absorptions lines that may be difficult to detect \citep[e.g.][]{burleigh99}. For now, we will assume the surface magnetic field is sufficiently low that component splitting is small compared with the WD line width. Therefore, the Lyman absorption equivalent widths should not be strongly impacted by Zeeman splitting \citep[e.g.][]{schmidt03}. Even these low fields strengths of a several MG, Zeeman splitting will tend to slightly broaden the line widths and result in underestimates of the surface temperatures based on observations of Lyman-$\alpha$. 

As seen in Figure~\ref{temperature}, the model spectra alone are a poor match the observed trough spectrum. The observed continuum is rather flat with no strong features while the DA models have a deep Lyman-$\alpha$ absorption. At surface temperatures of 13000$^\circ$K and hotter, the predicted WD model fluxes exceed the trough spectrum continuum and must be ruled out for our assumed mass and surface gravity.

\begin{figure}
    \centering
    \includegraphics[width=\columnwidth]{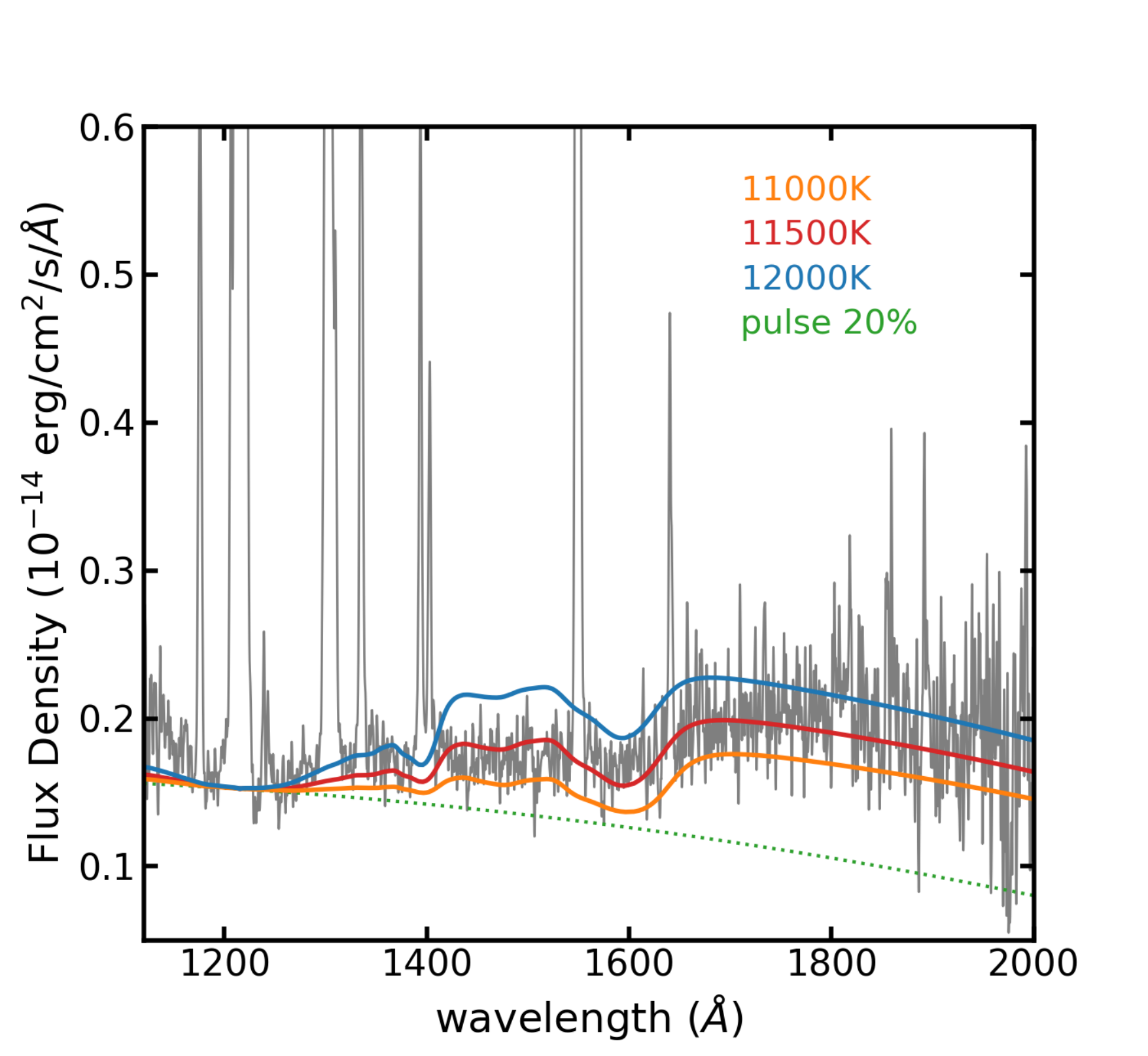}
    \caption{The trough spectrum of AR~Sco (light grey line) compared to WD models diluted by the pulsed spectrum (green dotted line). The pulsed spectrum has been scaled by a factor of 0.2 to match the observed spectrum at 1216~\AA . Diluted WD models with temperatures of 11000$^\circ$K, 11500$^\circ$K, 12000$^\circ$K, are shown as solid lines plotted over the observed spectrum. The weak dip observed at 1600~\AA\ is well matched by the models, but the models fall below the data shortward of 1200~\AA . (see \citet{lyutikov20})   }
    \label{dilute}
\end{figure}

The trough spectrum does show a shallow minimum around Lyman-$\alpha$, but the depth clearly does not match uncontaminated hydrogen-rich (DA) WD models. There is also a weak step in the continuum between 1600 and 1700~\AA, where there is a dip in the model spectra at temperatures less than 14000$^\circ$K.
This feature results from the quasi-molecular absorption by H$_2$ \citep{koester85}, and its depth is sensitive to temperature in this regime. In fact, the presence of the 1600~\AA\ feature in the observed spectrum implies that the WD must have a temperature of less than 14000$^\circ$K as the quasi-molecular absorption is too weak to detect above this temperature. 

A narrow dip in the model spectra at 1400~\AA\ is due to quasi-molecular absorption by H$^+_2$. No dip is seen the trough spectrum, but strong Si~IV emission interferes with the continuum at that wavelength.

The lack of a deep Lyman-$\alpha$ feature in the trough spectrum suggests, despite our attempts isolate it, that the light from the WD in AR~Sco remains diluted by another source. In creating the trough spectra, we were forced to include some the fading and rising sections of the pulsed emission centered around the minima, so we expect that the contamination is partly due to residual pulsed light. We can take the spectral shape of the pulses, add them to the WD models and attempt to match the trough spectrum. The WD models nearly go to zero flux at the bottom of the Lyman~$\alpha$ line, and this provides the scale to fix the pulse contamination. The sum of the scaled pulse spectrum and WD models around 11500$^\circ$K are compared with the observed trough spectrum in Figure~\ref{dilute}. The results are encouraging and suggest that the WD temperature is 11500$^\circ\pm 500^\circ$K. This is somewhat dependent on the assumed slope of the contaminating spectrum. Still, the observed dip around 1600~\AA\ is well matched by the sum of a WD model around 11500$^\circ$K added to an additional source with a blue continuum.

\subsection{The Pulsed Spectrum}

\begin{figure}
    \centering
    \includegraphics[width=\columnwidth]{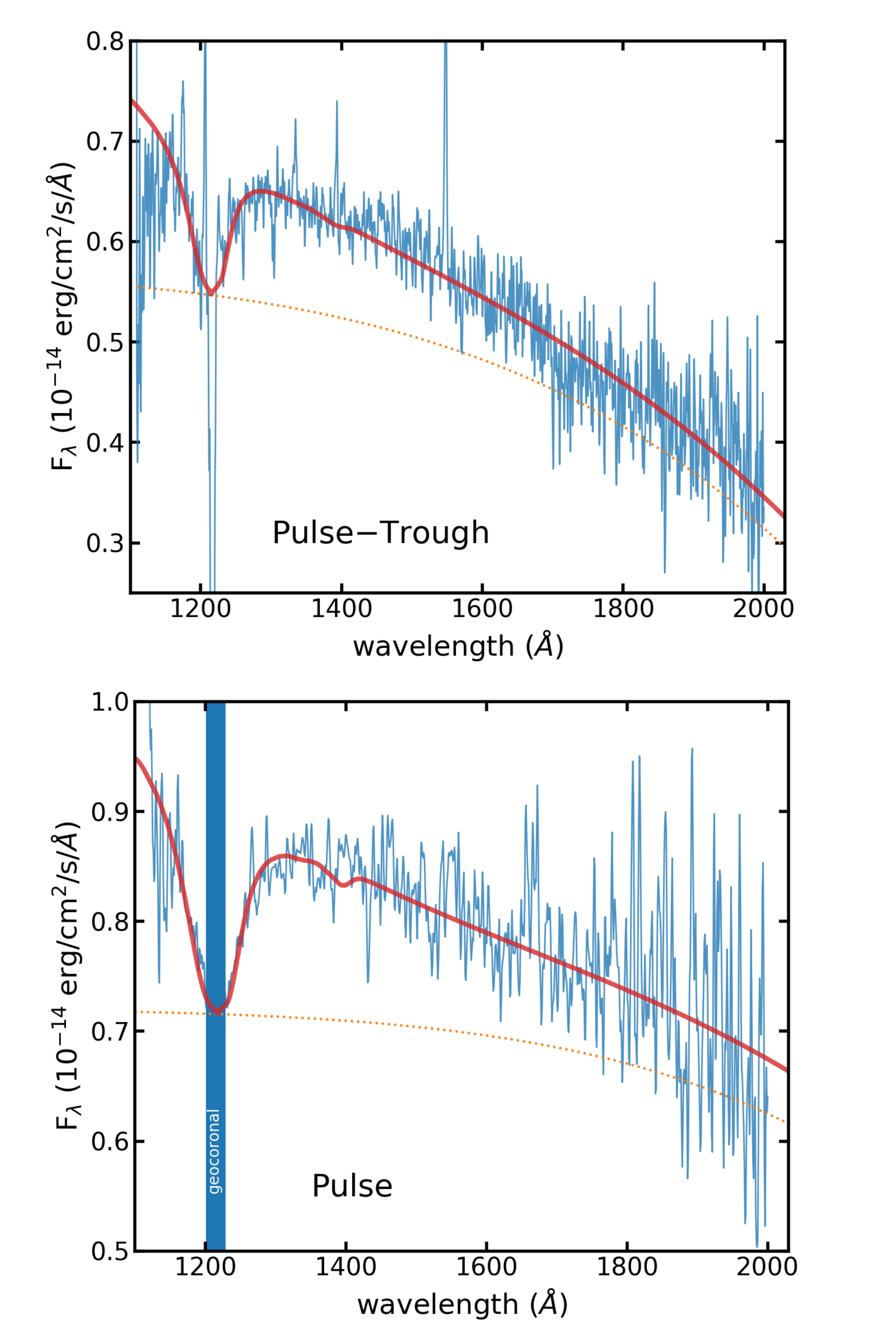}
    \caption{{\bf top:} The difference between the average pulse and trough spectra. The emission lines and geocoronal features are not completely subtracted leaving the spikes.  The broad absorption at Lyman-$\alpha$ is surprising because taking the difference between the pulse and trough spectra should eliminate the stellar contributions and leave only the spectrum of the pulsed emission. 
    The solid red line is a model made from the sum of a power-law and a 23000$^\circ$K WD model with $logg=8.5$ reduced in luminosity by a factor of 0.05. The dashed line shows the power-law added to the WD model to match the overall continuum. {\bf bottom:} The pulse spectrum with emission lines extracted. The thick vertical band indicates where the geocoronal Lyman-$\alpha$ dominates the flux. The solid red line is a model as described above, but using a 20000$^\circ$K WD model.  }
    \label{diff}
\end{figure}

The difference between the average pulse and trough spectra should contain pure synchrotron light with a minimum of contamination from stellar components. The difference spectrum is shown in the top panel of Figure~\ref{diff} and a strong blue continuum is apparent. The slope in $\nu F\nu$ is relatively flat as modelled by \citet{singh20} for the low-magnetic field synchrotron source (their synchrotron-1 region).

Surprisingly, there appears a dip in the continuum at the wavelength of Lyman-$\alpha$, when we would expect the synchrotron emission to be featureless. The broad absorption appears like the Lyman-$\alpha$ feature from the WD we expected to find in the trough spectrum. Taking the difference between the pulsed and trough spectra should have removed the stellar components if their contributions were constant over a WD spin period. The pulsed spectrum without subtracting the trough (lower panel of Figure~\ref{diff}) also shows an obvious broad Lyman-$\alpha$ absorption around the geocoronal emission. This is puzzling since the trough spectrum is four times fainter than the pulse around 1200~\AA\ and, the trough spectrum itself shows little evidence of Lyman-$\alpha$ from the WD. The strength of Lyman-$\alpha$ feature during the pulses implies that the WD thermal emission contributes about 20\%\ to the average pulse flux around 1200~\AA . In this data, the WD must be significantly brighter during the pulses than between pulses. 

For \citet{koester10} models with temperatures of 18000$^\circ$K$< T_{eff}< 30000^\circ$K, the full-width at half minimum (FWHM) of the Lyman-$\alpha$ absorption is approximately linear with surface temperature for a fixed surface gravity. For this range of temperatures, we normalized the model continua and fit the absorption with a Gaussian. This provides the relation:
$$T_{eff}=37170-221.7\times W \;\;\; ,$$ 
where $W$ is the measured FWHM in \AA . This linear relation reproduces the model temperatures with a standard deviation of 360$^\circ$K. We then measured the FWHM of the Lyman-$\alpha$ absorption in the pulsed spectra. The fit of the Gaussian from the $HST$ data is poorly constrained due to the geocoronal emission (and some Lyman-$\alpha$ emission from AR~Sco) that prevents a clear determination of the minimum, so uncertainties are large. For the pulse spectrum $W=55\pm 4$~\AA\ corresponding to a temperature of  25000$\pm 1000^\circ$K. The difference spectrum gives a somewhat lower FWHM of $W=39\pm 5$ corresponding to a temperature of 28500$\pm 1100$. The width of the Lyman-$\alpha$ absorption line suggests that the WD temperature during a pulse is a factor of two larger than seen during a trough. 

No WD spectral features other than Lyman-$\alpha$ are seen during the pulses, supporting the assertion that surface temperatures are in excess of 15000$^\circ$K where the quasi-molecular hydrogen features are no longer significant.

We also compared the \citet{koester10} WD models directly with the pulsed spectrum as shown in Figure~\ref{diff}.  We assume the flux below the Lyman-$\alpha$ minumum comes from a non-thermal power-law which is added to a WD spectral model. The best results are for WD temperatures of 20000$^\circ$K (pulse) and 23000$^\circ$K (pulse minus trough). These are slightly lower than the temperatures estimated using the absorption width alone, which may mean the assumption of a Gaussian line shape was not ideal. In these models, the WD fluxes need to be reduced by a factor of 16 (pulse) and 40 (pulse minus trough). The dilution suggests that the high temperatures do not cover the full WD surface area and may be localized hot regions or hot magnetic polar caps. Because we do not have a good picture of their temperature structure or extent, We will refer to these regions as ``hotspots''.

The presence of a clear, single, Lyman-$\alpha$ absorption that appears to be symmetric about 1215~\AA\ implies that the surface magnetic field on the WD is less than $\lesssim$100~MG. Zeeman splitting becomes quite apparent above this field strength as seen in AR~UMa \citep{gaensicke01}. This is a fairly conservative limit, but the geocoronal emission in the AR~Sco data could hide subtle effects of Zeeman splitting at lower field strengths.

\subsection{The Visibility of the Hotspots}

The apparent hotspots on the WD inferred from these data are seen primarily during the beat pulses and are not prominent during the inter-pulse periods. However, we expect the temperature variations on the WD to be seen at its spin period, while the pulses are observed at the beat period. The combination of beat and spin pulses was modelled by \citet{stiller18}, who showed how the interference of the beat and spin modulations shapes some of the properties of the optical light curve. They note that constructive interference between these two signals occurs at orbital phase $0.25 < \phi < 0.35$ and at half an orbit later. Destructive interference is greatest near $\phi\sim 0.05$ and $\phi\sim 0.55$. Similar orbital phasing between these signals is seen in the FUV light curve. As a result, we would expect the hotspots to be contributing to the trough light at certain orbital phases, while during other parts of the orbit, they would coincide with the synchrotron pulses. Yet evidence for the $>$20000$^\circ$K temperatures is seen in the pulse spectrum and not in the trough spectrum. 

There are two reasons for the hotspots to have been predominately visible during the pulses. First, the binary orbital sampling by $HST$ was not uniform. The observation consists of five spacecraft orbits each taking 95 minutes, with gaps caused by Earth occultations that last 45 minutes. This resulted in the data skipping binary orbital phases near $\phi\approx 0.1$ and $\phi\approx 0.6$ while phases around $\phi\approx 0.35$ and $\phi\approx 0.8$ were covered twice. So the phases when the spin and beat destructively interfere were poorly sampled and the constructive sections overly represented in the data. By chance, this meant the beat pulses tended to contain spin peaks caused by the hotspots. Conversely, the cool regions of the WD tended to be visible during the troughs.

The second bias against including the hotspots in the inter-pulse regions is that when the spin was out of phase with the beat, the troughs tended to be too bright to include in the average. As discussed in Section~2, troughs were defined as periods with fluxes fainter than about 2$\times 10^{-15}$ \flunits , while the peaks of the spin modulation would push the light curve in the inter-pulse regions above that limit. Thus, the trough spectra were biased against including the hotspots.

The Lomb-Scargle power spectrum of the FUV light curve \citep{marsh16} shows that there is almost no power at the fundamental WD spin frequency, but the very strong peak at twice the spin frequency implies that the temperature of the two hotspots is very similar. From the power spectrum, the average amplitude of the spin peaks is about 20\%\ to 30\%\ of the primary beat pulse amplitudes. This is consistent with what we see in the spectra where the Lyman-$\alpha$ depth is suggests the hotspots add 20\%\ to the pulse flux.

In the optical, there is little contribution of the hotspots to the observed spin modulation. Extrapolating a 23000$^\circ$K WD model at the distance of AR~Sco and diluted by a factor of 40 implies a $SDSS-r$-band brightness for the hotspot contribution to be only $\approx 21$~mag. The $SDSS-r$ brightness of AR~Sco ranges between 14~mag and 16~mag \citep{gaibor20}. Thus, the amplitude of the hotspot modulation in the optical would amount to a few percent of the total flux and be lost in the other variations in the system. Instead, the linear polarization at the spin frequency observed by \citet{potter18} implies some non-thermal source of emission is important to the spin modulation at optical wavelengths.

\subsection{Spin Color Variability}

The spectral changes of the WD over the expected temperature range is particularly dramatic in the FUV. For the relatively cool 11500$^\circ$K the quasi-molecular absorption severely decreases the flux short-ward of 1600~\AA\ (see Figure~\ref{temperature}). In contrast, the $>$20000$^\circ$K spectrum is quite blue and unaffected by quasi-molecules. Splitting the FUV spectra into two bands divided at 1600~\AA, we predict a large color modulation as the WD rotates.

For the AR~Sco FUV data, we construct a blue bandpass running between 1150~\AA\ and 1500~\AA\ and a red bandpass between 1600~\AA\ and 2000~\AA . We convert the average flux in each bandpass to a magnitude and construct a color in the standard convention by subtracting the red magnitude from the blue. The resulting color curve shows a large modulation at a period of around one minute.

Using the \citet{koester10} WD models scaled to the distance of AR~Sco, We can estimate the expected color modulation from the 11500$^\circ$K rotating WD with hotspots around 23000$^\circ$K. The median flux level for the $HST/COS$ light curve is 3.0$\times 10^{-15}$ \flunits\ which we add to the model spectra before calculating the colors using the same prescription as for the real data. The difference in color between the uniform WD and the WD with hotspots amounts to 0.23~mag, so we expect the color modulation of the spinning WD to have approximately this peak-to-peak amplitude.

To separate the beat and spin color modulations, we calculate the color Lomb-Scargle power spectrum in Figure~\ref{color_pow} and compare it to the flux power spectrum. Twice the spin frequency, $2\omega$, shows the highest peak and this corresponds to a color amplitude of 0.30$\pm 0.02$~mag. This is fairly consistent with what we expect from the hotspot model. The very low power at the native spin frequencies suggest that the double peaks viewed each cycle are nearly equal in amplitude.

Double the beat frequency, $2(\omega -\Omega)$, also has significant power corresponding to an amplitude of 0.20$\pm 0.02$~mag. This color modulation at may correspond to variations in the FUV power-law slope of the synchrotron emission during both the primary and secondary beat pulses.

\section{Discussion}

\subsection{Comparison to AE Aqr}

Our analysis suggests that during the pulsed emission, the WD in AR~Sco has a surface temperature above 20000$^\circ$K, while during the inter-pulse segments the surface temperature is around 11500$^\circ$K. This can be explained with a hotspot near the magnetic poles combined with a cooler equatorial band. Accreting magnetic WDs such as in the proto-type AM~Her have been observed to have hotspots at their magnetic poles \citep{gaensicke06}. Hotspots also explain the fast brightness variations in the propeller-state IP AE~Aqr \citep{eracleous94}. The 33s UV variability in AE~Aqr is the consequence of temperature variations on the surface of the spinning WD in the system. The spectrum of the pulse peaks shows a broad Lyman-$\alpha$ absorption and a continuum consistent with a WD temperature of 26000$^\circ$K. \citet{eracleous94} found an excellent fit to the continuum variability by assuming a minimum temperature of 14000$^\circ$K. These are very close to the range of temperatures we have estimated for AR~Sco. In AE~Aqr, the UV light curve on short time-scales is completely dominated by the temperature variations on the WD. 

For AR~Sco, the UV variations result from the combination of the WD surface hotspots and non-thermal pulsed emission. The WD temperature dipole variations contribute 20\%\ to the pulses around 1200~\AA\ and a decreasing fraction at longer wavelengths. As discussed in the previous section, there are times when the hotspots are out of phase with the pulses, but the gaps in the existing $COS/HST$ data and the way we extracted the troughs tended to avoid times when the poles would be more cleanly separable from the pulses. 

The temperature structure on the WD in AE~Aqr is similar to what we expect in AR~Sco. This color change on the spin period should be directly visible on AE~Aqr since it does not have a large beat pulse confusing the spin variations. Indeed, \citet{eracleous94} divided the FUV spectra into several wavelength bins and found the highest amplitude for the spin modulation in the FUV was at the 1340~\AA\ and 1450~\AA\ bins and lowest at 2000~\AA . From their plots we find a color amplitude of between 0.2 and 0.4~mag. In the previous section we showed a color amplitude of 0.30~mag for AR~Sco, in excellent agreement with that for AE~Aqr.

For AE~Aqr, the FUV flux modulation at the spin period shows two unequal amplitude peaks suggesting that the magnetic axis and viewing angle conspire to dim our view of one of the hotspots  \citep{eracleous94}. In contrast, the FUV flux power spectrum for AR~Sco implies that the spin modulation is very symmetric so that the magnetic axis is nearly perpendicular to the spin axis and that the hotspots in AR~Sco are very close to the WD equator.

\begin{figure}
    \centering
    \includegraphics[width=\columnwidth]{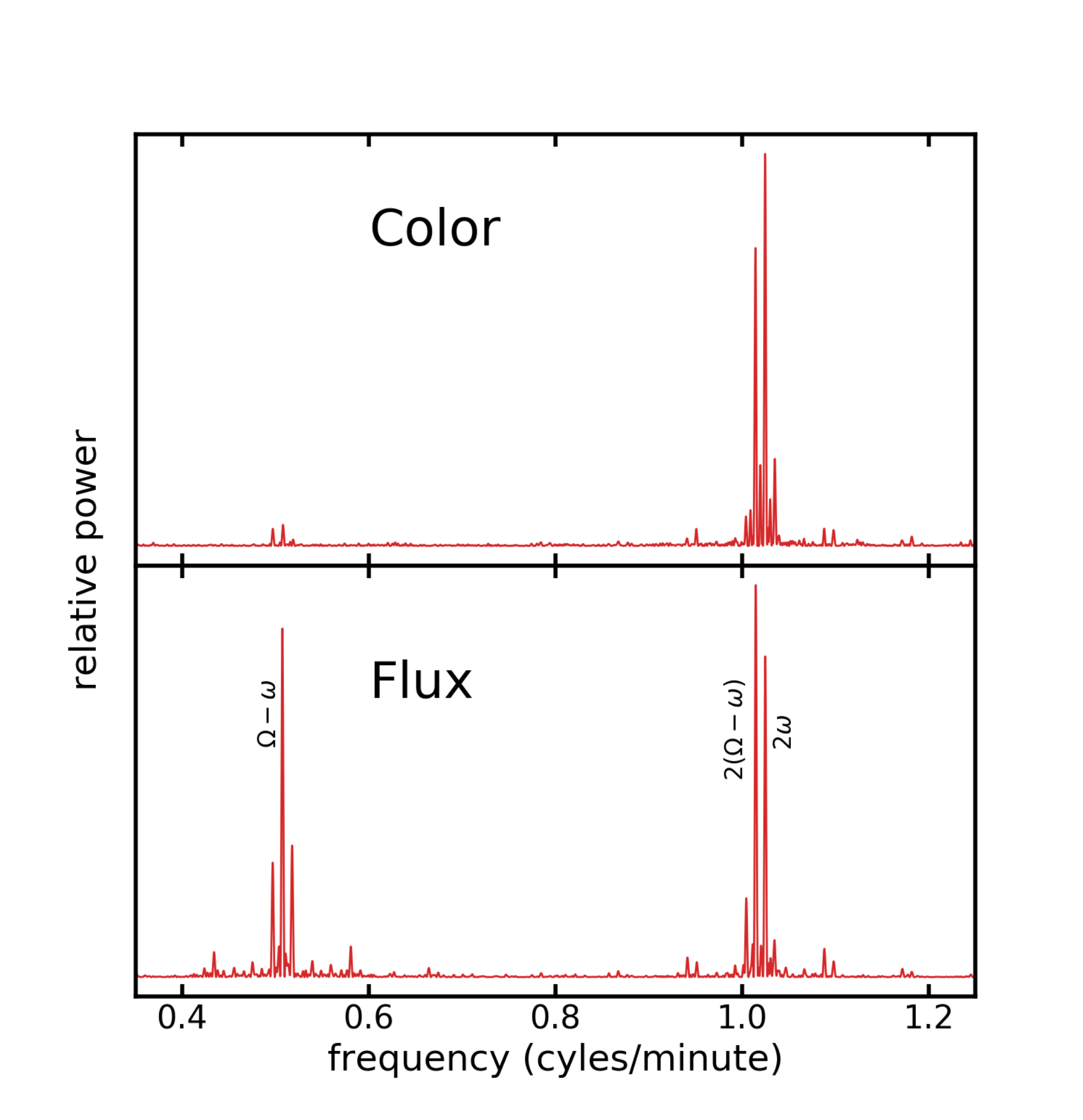}
    \caption{{\bf top:} The power spectrum of the color curve created by dividing the FUV spectra into a short wavelength band (1150-1500~\AA ) and a long wavelength band (1600-2000~\AA ). The strongest peak corresponds to twice the spin frequency and it has an amplitude of 0.30~mag.  {\bf bottom:} The power spectrum of the total FUV flux. The weak peaks near each of the major peaks result from the way $HST$ sampled the light curve. }
    \label{color_pow}
\end{figure}

\subsection{Origin of the Hotspots}

Some isolated magnetic WDs show low-amplitude oscillations at their spin period due to spots on their surface \citep[e.g.][]{hoard18}. It has been suggested that brightness modulations can be caused by strong magnetic fields or localized abundance enhancements that result in surface temperature variations \citep{holberg11}. While these modulations are only a few percent in the optical, the temperature variations may have a larger impact in the FUV. In mass-transfer binary systems, such as AE~Aqr, direct accretion on to the magnetic poles is also a possible cause of WD hotspots.  \citet{marsh16} constrained the WD accretion rate in AR~Sco to be less than $1.3\times 10^{-11}$~M$_\odot$~yr$^{-1}$ and pointed out that accretion near that rate would result in broad emission lines distinct from the narrow lines observed on the secondary star. Accretion at a lower rate has not been ruled out and could be a source of some heating. Because of the unique nature of AR~Sco, we present two new models to explain the apparent temperature gradients on its WD.

\subsubsection{Hot spots on the WD due to synchrotron heating?}
\label{sc:sh}

\citet{potter18} described a  model in which  relativistic electrons  following the field lines 
produce synchrotron radiation beamed in the direction of the WD. The opening angle for the beam was quite large, so that only a small fraction of the energy would irradiate the surface of a WD at its magnetic poles.

Assuming equilibrium, the thermal energy emitted by the hotspots should match the synchrotron energy absorbed on the poles. In the previous section, we estimated that at 23000$^\circ$K, a hotspot would cover 2.5\%\ of the WD surface. This hotspot is above the typical surface temperature of 11500$^\circ$K and generates an excess luminosity of 1.7$\times 10^{30}$ erg~s$^{-1}$ that must be replenished through heating. \citet{marsh16} estimated that the synchrotron production in AR~Sco is $L_{syn}=1.3\times 10^{32}\;$~erg~s$^{-1}$, meaning that only 1\%\ of the generated energy needs to be absorbed at a magnetic pole to maintain the hotspot temperature.

\citet{bl2020} has modelled the particle motions in the WD magnetosphere and included effects of radiative damping  and adiabatic mirroring. These results suggest that synchrotron emission from energetic electrons approaching the magnetic poles is unlikely to produce localized heated region on the WD. As non-thermal electrons move from their acceleration zone along magnetic field lines towards the poles, they increase their pitch angle $\alpha$ (see \citet{kpa15}). As a result, at the mirror point, where the emissivity is maximal, the emissivity diagram is very wide, predominately in the direction  normal to the magnetic field lines. Thus, this thin cone of synchrotron emission will mostly miss the WD surface. 
For some combinations of initial pitch angles and lepton Lorentz factors, the emission cone will touch the WD surface, but it would irradiate nearly the entire WD hemisphere.  

Thus, emission from energetic electrons trapped in the WD magnetosphere is unlikely to be the source of localized heating on the WD surface.

\subsubsection{Baryonic Bombardment}
\label{sc:hb}

Another possible source of heating of the polar caps may be
by the leaking of protons trapped on closed field lines of the WD’s magnetosphere.
Trapping occurs through turbulent scattering, also known as radial diffusion in the case of the Earth magnetosphere. One of the most important processes will be magnetic reflection due to the conservation of the first adiabatic invariant. As a result, only a small fraction of baryons can reach the WD surface $\delta \approx 1/8 (B_{WD}/B_{A})^{-1} \approx 1/8 (R_{WD}/R_{A})^3 \approx 10^{-6}$ for each cycle of magnetospheric oscillations. 

In the model suggested by \citet{lyutikov20} the matter flow from the red dwarf secondary interacts with the WD magnetosphere at a radius of about $R_A\approx2\times10^{10}$~cm.
At this point electrons and baryons are accelerated in magnetic reconnection events. As electrons propagate within the magnetosphere of the WD, they experience strong radiative cooling, losing their energy to synchrotron emission. As baryons do not emit efficiently, their energy is conserved.

The captured baryons will oscillate on the magnetic field lines between the magnetic poles, forming analogs of Earth's van Allen radiation belts. Scattering on turbulent fluctuations will lead to a changing pitch angle. After approximately  $N\sim 10^3$ oscillations they will be scattered into the loss cone and will hit the WD.

Another process limiting the baryons' lifetime in the WD magnetosphere is the escape from the magnetosphere at a characteristic time of $t_{es}\sim 2\pi \eta/\omega$, here $\omega$ is WD angular velocity and $\eta\sim 1$ is the factor which takes into account probability for non-thermal (NT) particle to escape \footnote{Even for $10^3$ oscillations the synchrotron cooling time is negligible for baryons with energy bellow 1~TeV.}, so the number of oscillations per baryon is $N\sim t_{es} c/\pi R_A$.

The polar cap heating rate can be estimated as $L_{NT}\delta N \sim \mbox{few}\times10^{30}$~erg/s, where $L_{NT}\gtrsim L_{syn}$ is proton acceleration rate. This heating rate is close to the observed hotspot luminosity. We note that the randomization from many oscillations through the magnetosphere will result in approximately equal amounts of energy deposited on the two poles.  Indeed, an appealing feature of this mechanism is that it naturally explains how the observed hotspots have such similar temperatures (see Figure~\ref{color_pow}). 


\subsection{Implications for theoretical models of AR Sco}

\citet{lyutikov20} have proposed a mechanism to explain the observed characteristics of AR~Sco that does not require accretion on to the WD. They predict that neutral gas leaving the secondary will reach into the WD magnetosphere before being ionized and ejected through the propeller mechanism. To do this, they estimate that the WD requires a surface temperature of at least 12000$^\circ$K for sufficient ionization of the secondary's gas. Here, we find that during the time between pulses the WD surface temperature is consistent with their prediction within our range of uncertainty. 

However, there are several results here that may complicate \citet{lyutikov20} interpretation. For example, the magnetic poles of the WD appear to have temperatures in excess of 20000$^\circ$ which would ionize the gas coming from the secondary before it reaches the inner magnetosphere of the WD. Our analysis of the trough spectrum shows that the inner face of the secondary may already be at a temperature of $\approx$15000$^\circ$K and likely to have a high ionization fraction before heading toward the WD. Indeed, optical spectra have already found HeII 4686~\AA\ emission located on the secondary star's inner face \citep{garnavich19} implying heating through irradiation or magnetic interaction.

From these spectra, we cannot directly estimate the magnetic field strength on the WD surface. All we can do is note that there is no clear Zeeman splitting of the Lyman-$\alpha$ absorption which suggests that the fields are less than about 100~MG \citep{gaensicke01}. While rather high, this limit puts pressure on models where the WD magnetic field is in the 200~G range at the red dwarf. Such models require WD surface fields of $>200$~MG \citep[e.g.][]{takata18, garnavich19}. Likewise, our constraint challenges a scenario proposed in \citet{buckley17} wherein the WD field strength could be as high\footnote{The original estimate in \citet{buckley17} was $\sim$500~MG, but per their Eq.~1, this value needs to be doubled due to improved measurements of the spin-down power from subsequent studies \citep{stiller18, gaibor20}.} as $\sim$1000~MG.

\section{Conclusion}

In order to constrain the temperature of the WD in AR~Sco, we reanalyzed the FUV spectral data obtained with $COS$ on $HST$ in 2016. The average spectrum obtained over two binary orbits does not show clear evidence of emission from the WD in the system. We divide up the observations in time to build up an average ``trough'' spectrum from the minima between pulses. We further sub-divide the trough spectra by orbital phase to minimize emission sources that contaminate the WD signal.

From the trough spectra averaged around inferior conjunction, we find plausible evidence for a quasi-molecular hydrogen absorption band and flux constraints that imply a WD surface temperature of 11500$\pm 500^\circ$K.

The trough spectra obtained around superior conjunction shows broad excess emission peaking around 1900~\AA . This emission corresponds to the peak of the orbital modulation and it is likely coming from the irradiated face of the secondary star that is best viewed around orbital phase 0.5. The spectral peak implies a hot region on the irradiated face of the secondary at about 15000$^\circ$K. 

An average spectrum of the pulsed emission shows a strong Lyman-$\alpha$ absorption that is not obvious between pulses. The pulsed spectrum is well fit by a power-law pulse continuum added to a WD spectrum with a temperature of 23000$\pm 3000^\circ$K. The WD flux contributes 20\%\ to the total pulsed light around 1200~\AA . We conclude that hotspots are present on the WD and visible primarily during the pulses.

Magnetic WDs in polars often have hotspots due to direct accretion of gas lost by their companion stars. Some direct accretion on to the WD in AR~Sco has not been ruled out, and may be a source of the observed localized heating. Another possible heating source comes from the \citet{potter18} model that has the synchrotron beams approximately aimed in the direction of the WD magnetic poles, suggesting that the hotspots result from absorption of a fraction of the synchrotron energy. More detailed modelling of electron emission in the WD magnetosphere \citep{bl2020} described in section \ref{sc:sh} suggests that the synchrotron beams are unable to provide the desired localized heating. Here, we present a new scenario where non-thermal proton bombardment deposits energy at the WD magnetic poles. These trapped protons make many oscillations between the poles that naturally provides an explanation for the similar temperatures of the two observed hotspots.

The FUV spectra do not strongly constrain the surface magnetic field on the WD. Geocoronal emission makes it difficult to detect details in profile of the Lyman-$\alpha$ line. From this data we only conclude that there is no strong Zeeman splitting visible and constrain the field to less than 100~MG. Still, this means the WD field at the surface of the red dwarf is less than about 50~G.

Our results support the \citet{lyutikov20} model in that the WD temperature is sufficient to ionize gas from the red dwarf as it approaches the inner magnetosphere. In fact, the temperature of the WD hotspots may be so high that gas is significantly ionized close to the secondary. We see evidence for 15000$^\circ$K gas on the irradiated face of the secondary, implying a fairly high ionized fraction near the $L_1$ point. While this does not rule out the \citet{lyutikov20} model, it does suggest that the interaction mechanism between these two stars requires further investigation.

\begin{acknowledgements}

This research is based on observations made with the NASA/ESA Hubble Space Telescope obtained from the Space Telescope Science Institute, which is operated by the Association of Universities for Research in Astronomy, Inc., under NASA contract NAS 5–26555. These observations are associated with program GO14470, PI: Gaensicke. This work had been supported by DoE grant DE-SC0016369,
NASA grant 80NSSC17K0757 and NSF grants {1903332 and 1908590}.
ML would like to thank organizers and participants of the conference ``Compact White Dwarf Binaries'' for enlightening discussions.

\end{acknowledgements}
    

\end{document}